\documentclass[fleqn,usenatbib]{mnras}
\usepackage{newtxtext,newtxmath}

\usepackage[T1]{fontenc}
\usepackage{ae,aecompl}

\usepackage[version=4]{mhchem}    
\usepackage{graphicx}	
\usepackage{amsmath}	

\title[Temperature inversions: Super-Earths]{Temperature inversions on hot super-Earths: the case of CN in nitrogen-rich atmospheres}

\author[M.Zilinskas et al.]{
Mantas Zilinskas,$^{1}$\thanks{E-mail: zilinskas@strw.leidenuniv.nl}
Yamila Miguel$^{1}$
Yipeng Lyu$^{1}$
Morris Bax$^{1}$
\\
$^{1}$Leiden Observatory, Leiden University, Niels Bohrweg 2, 2333CA Leiden, The Netherlands\\
\\
}

\date{Accepted XXX, Received YYY; in original form ZZZ}

\pubyear{2020}

\begin{document}
\label{firstpage}
\pagerange{\pageref{firstpage}--\pageref{lastpage}}
\maketitle

\begin{abstract}
We show that in extremely irradiated atmospheres of hot super-Earths shortwave absorption of \ce{CN} can cause strong temperature inversions. We base this study on previous observations of 55 Cancri e, which lead us to believe that ultra-short-period super-Earths can sustain volatile atmospheres, rich in nitrogen and/or carbon. We compute our model atmospheres in a radiative-convective equilibrium for a variety of nitrogen-rich cases and orbital parameters. We demonstrate the effects caused by thermal inversions on the chemistry and compute low resolution synthetic emission spectra for a range of 0.5 - 28 $\micron$. Our results indicate that due to shortwave absorption of \ce{CN}, atmospheres with temperatures above 2000 K and C/O $\geq$ 1.0 are prone to thermal inversions. \ce{CN} is one of the few molecules that is extremely stable at large temperatures occurring on the day side of short period super-Earths. The emission spectrum of such atmospheres will differ substantially from non-inverted cases. In the case of inversions, absorption features become inverted, showing higher than expected flux. We propose that inversions in hot atmospheres should be the expected norm. Hot super-Earths are some of the most extreme natural laboratories for testing predictions of atmospheric chemistry and structure. They are frequently occurring, bright in emission and have short orbital periods. All these factors make them perfect candidates to be observed with JWST and ARIEL missions. 
\end{abstract}




\section{Introduction}
\label{Introduction}



Phase curve observations of 55 Cancri e \citep{Demory_2016b}, a nearby ultra-short-period (USP) (P < 1 day) super-Earth, revealed an eastwards shifted hotspot away from the planet's substellar point, with a high antistellar temperature of $\sim$1400 K and a significant, $\sim$1300 K day-night temperature contrast. The high antistellar temperature can only be sustained by heat transport from the day side, however, because the day-night temperature difference is also high, atmospheric heat redistribution between the two hemispheres cannot be large. The heat can be transported either on the planet's surface or in its atmosphere. \citet{Kite_2016} argued that the surface magma currents alone would be insufficient to sustain the observed nightside temperature, calling for the need of a substantial atmosphere. Further modelling studies by \citet{Hammond_2017,Angelo_2017,Bourrier_2018} have all suggested a thick atmosphere as an explanation for the observed features.

While the bulk component of 55 Cancri e's atmosphere has been subject to some speculation, models and observations provided by  \citet{Bourrier_2018} show that a scenario of a light-weight, hydrogen-rich atmosphere is extremely unlikely. The planet's close proximity to the host star also implies intense atmospheric erosion, which would allow hydrogen to escape over a period of time much shorter than its current age. The planet's density, orbital parameters and required  heat redistribution efficiency is well in agreement with the atmosphere being composed of high-mean-molecular-weight gas, as suggested by \citet{Demory_2016b,Angelo_2017,Bourrier_2018}. No clear detection of \ce{H2} or \ce{H2O} in the planets atmosphere \citep{Ehrenreich_2012,Esteves_2017}, and more recent, high-resolution observations \citep{Jindal_2020} further support the case of a high-mean-molecular-weight atmosphere.

The derived heat redistribution coefficient by \citet{Angelo_2017} indicate that the atmosphere is 1 - 2 bar. Assuming no thermal inversion occurs in the atmosphere, the dayside temperature of the planet is consistent with the expected equilibrium surface temperature, indicating that the Spitzer IRAC 4.5 \micron{} bandpass probes deep into the atmosphere \citep{Demory_2016b,Angelo_2017}. With current opacity data this favours atmospheres composed mostly of either \ce{N2} or even \ce{CO}. Atmospheres with an abundance of \ce{H2O} or \ce{CO2} would have large absorption in the bandpass and thus are dismissed.

While the nature of USP super-Earth atmospheres is still highly unknown, from observations in our own solar system we know that nitrogen is a rather common denominator of rocky body atmospheres \citep{Gladstone_2016, Wong_2017,Lammer_2018,Grenfell_2020}. In geological studies, nitrogen is also found to be one of the common gases to be accreted and outgassed during the evolution of rocky planets \citep{Wordsworth_2016,Lammer_2018,Lammer_2019}. Therefore, in this work, we make an assumption that USP super-Earths, such as 55 Cancri e, can evolve to have \ce{N2}-rich volatile atmospheres.

In previous works, \citep{Miguel_2019,ZILINSKAS_2020} have explored possible nitrogen chemistry on 55 Cancri e and found that with just a small amount of hydrogen in the atmosphere, hydrocarbons such as \ce{HCN} would show strong absorption in emission through secondary eclipse. It was also suggested that in the absence of large amounts of hydrogen, which could occur if the planet experienced severe hydrogen escape throughout its history, \ce{CN} would be the dominant nitrogen carrier after \ce{N2}. \ce{CN} molecule is of great interest because it is a significant shortwave absorber. Shortwave absorption can lead to temperature inversions \citep{Hubeny_2003,Fortney_2008,Haynes_2015}, consequently affecting the chemistry and spectral features. While \ce{CN} is very reactive, it is also one of the few molecular composites that are stable at very large atmospheric temperatures (up to $\sim{}$4000 K), characteristic of short period planets.

In this paper, we aim to explore the possibility and consequential effects of temperature inversions caused by \ce{CN} in nitrogen-rich atmospheres of USP super-Earths. While we take the parameters 55 Cancri e as an example our intention is not to predict its exact spectral features. Instead, by using numerical radiative transfer and chemical kinetics models, we explore a variety of possible nitrogen-rich atmospheric compositions at different orbital distances, hence different temperature ranges. Since we expect close-in planets to experience severe hydrogen depletion over its evolution, we also simulate a range of compositions with different abundances of hydrogen.

Currently, possible  atmospheric compositions of USP super-Earth are still very ambiguous. The upcoming missions of \textit{James Webb Space Telescope} (\textit{JWST}) and \textit{ARIEL} will allow us to probe a large spectral range, giving more insight in what the bulk atmospheric constituents might be, however, correct interpretation requires good understanding of the underlying theory, such as the occurrence of temperature inversions, which can have a substantial effect on the chemistry, and consequently emission spectroscopy. 

This paper is arranged as follows. In Section \ref{sec:methods} we describe our methodology, outlining the used numerical models, atmospheric compositions and system parameters. We present the resulting temperature profiles, chemical compositions and emission spectra in Section \ref{sec:results}. We discuss the importance and underlying issues of our key findings in Section \ref{sec:discussion}. Section \ref{sec:conclusion} contains the summary and the conclusion of our study.

\section{Methods}
\label{sec:methods}
\subsection{Planetary parameters and initial compositions}
\label{meth:planetparams}
The goal of this study is to examine the possibility of temperature inversions in atmospheres where nitrogen is the dominant component. Since, observations of 55 Cancri e strongly suggest a nitrogen atmosphere, we take its physical parameters as our starting point. For the radius and mass we use $r = 1.897$ R${_\oplus}$ and $m = 8.59$ M${_\oplus}$, respectively \citep{Bourrier_2018}. Following \citet{Angelo_2017} we assume a 1.4 bar surface pressure. As in \citep{Miguel_2019,ZILINSKAS_2020} we take our initial atmospheric composition to be that of Titan's, which, contains 96$\%$ nitrogen, making it the most extreme example of a known nitrogen-dominated atmosphere. We only consider atmospheres composed of \ce{H}, \ce{N}, \ce{C}, and \ce{O} elements, neglecting the possibility heavier metals outgassed from the surface. Since \ce{CN} formation is heavily dictated by the C/O ratio, we vary, both, carbon and oxygen abundances to create compositions with C/O ratios ranging from 0.5 to much larger than unity. Because depletion of hydrogen could effect the amount of \ce{CN} produced, for each composition we additionally explore hydrogen mass fractions ranging from $3 \times 10^{-8}$ to $3 \times 10^{-2}$. To exaggerate the effect of hydrogen loss, the amount of lost hydrogen is always replaced by additional nitrogen. In order to highlight spectral differences caused by inversions in nitrogen-dominated atmospheres, we also explore three cases where the main constituent in the atmosphere is \ce{CO} or \ce{H2}. The compositions of the atmospheres that we adopt in this paper are displayed in Table \ref{tab:comps}.

\begin{table}
 \caption{Initial atmospheric elemental compositions used in our study. For each named type we explored four different compositions with varied hydrogen mass fractions (as shown for the `Titan' atmosphere case ).}
 \label{tab:comps}
 \begin{tabular}{llllll}
 \hline\noalign{\smallskip}
  Type & Nitrogen & Hydrogen & Carbon & Oxygen & C/O\\
  \noalign{\smallskip}\hline\noalign{\smallskip}
  Titan & 9.63E-01 & 3.00E-02 & 7.00E-03 & 2.50E-05 & 280\\
  & 9.93E-01 & 3.00E-04 & 7.00E-03 & 2.50E-05 & 280\\
  & 9.93E-01 & 3.00E-06 & 7.00E-03 & 2.50E-05 & 280\\
  & 9.93E-01 & 3.00E-08 & 7.00E-03 & 2.50E-05 & 280\\

  N2 (2.5)& 9.63E-01 & 3.00E-02 & 5.02E-03 & 2.01E-03 & 2.5\\
  & ... & ... & ... & ... & ...\\

  N2 (1.0)& 9.63E-01 & 3.00E-02 & 3.51E-03 & 3.51E-03 & 1.0\\
  & ... & ... & ... & ... & ...\\
 
  N2 (0.5)& 9.63E-01 & 3.00E-02 & 2.34E-03 & 4.68E-03 & 0.5\\
  & ... & ... & ... & ... & ...\\
  \hline 
  CO (1.0) & 7.03E-03 & 3.00E-02 & 4.81E-01 & 4.81E-01 & 1.0\\
  H2 (2.5) & 3.00E-02 & 9.53E-01 & 7.00E-03 & 2.80E-03 & 2.5\\
  H2 (0.55) & 3.00E-02 & 7.23E-01 & 7.23E-02 & 1.32E-01 & 0.55\\
  \noalign{\smallskip}\hline 
 \end{tabular}
\end{table}

Temperature is an important factor in atmospheric chemistry. The variation of the orbital distance impacts the irradiation that the planet experiences, which will significantly vary the atmospheric temperature profile and thus the resulting chemistry. For all our compositions we vary the orbital radius ranging from 0.007 to 0.06 AU which equates to a T\textsubscript{Peff} range of 1274 K to 3730 K. In total we explore 5 different temperatures for a total of 19 initial compositions. We found this range to be sufficient to cover the possible atmospheric structure diversity for hot super-Earths. Note, T\textsubscript{Peff} here represents a surface temperature if the planet were to emit purely as a blackbody.

\subsection{Temperature profiles}
\label{meth:temperature}

We numerically calculate our temperature profiles by computing the radiative-convective equilibrium with the open source radiative transfer code \textsc{HELIOS}\footnote{https://github.com/exoclime/HELIOS} \citep{Malik_2017,Malik_2019}. Besides planetary system parameters and stellar spectrum, radiative transfer codes require opacity tables as input. Our opacity k-tables are constructed using pre-calculated opacities obtained from the opacity database\footnote{http://opacity.world} \citep{Grimm_2015}. As absorbers we include: \ce{C2H2}, \ce{C2H4}, \ce{CH}, \ce{CH3}, \ce{CH4}, \ce{CN}, \ce{CO}, \ce{CO2}, \ce{H2O}, \ce{HCN}, \ce{NH}, \ce{NH3}, \ce{NO}, \ce{O} and \ce{OH}. For each of our compositions listed in Table \ref{tab:comps} these are weighted according to molecular abundances obtained using a thermochemical equilibrium model \textsc{FastChem}\footnote{https://github.com/exoclime/FastChem} \citep{Stock_2018}. For the k-table sampling wavelength resolution we use $\lambda/\Delta\lambda = 1000$, accounting for a range between 0.06 and 200 $\mu$m.
 
For stellar irradiation, instead of using a blackbody, which tends to overestimate shortwave radiation, we use a synthetic stellar spectra of T\textsubscript{eff} = 5250 K with coadded UV observations of a similar star to 55 Cancri  \citep{Kurucz_1979,Rugheimer_2013,ZILINSKAS_2020}. For the stellar radius we take 0.954 solar radii \citep{Bourrier_2018}. The temperature profiles are calculated assuming day side redistribution ($f = 2/3$) with the diffusivity factor set to 2, surface albedo to 0 and internal temperature to 0.

\subsection{Chemical evolution}
\label{meth:chemistry}
In order to investigate the effects of temperature inversions on chemistry and consequently emission spectra, we evolve our atmospheres using the calculated temperature profiles from equilibrium compositions with a chemical kinetics model \textsc{VULCAN}\footnote{https://github.com/exoclime/VULCAN} \citep{Tsai_2017}. Note that our temperature profiles are calculated using equilibrium chemistry, while subsequent chemistry is explored with chemical kinetics. We use a validated N-C-H-O network consisting of over 600 reactions (including reversed) and over 50 molecular species. The network is available as part of the \textsc{VULCAN} package.
As disequilibrium processes we include the effects of photochemistry, atmospheric mixing (defined with K\textsubscript{zz}) and molecular diffusion. Our photochemical rates are calculated using
\begin{equation}
    k(z)=\int q(\lambda) \sigma_{d}(\lambda) J(\lambda, z) d \lambda
    \label{eq:photodissociation}
\end{equation}
where, $q(\lambda)$ represents the quantum yield, defined as the rate of occurrence of certain photodissociation products per absorbed photon, $\sigma_{d}(\lambda)$ - absorption cross-section, $J(\lambda, z)$ - calculated actinic flux from, both, direct stellar beam and scattered radiation contributions using a two-stream approximation \citet{Malik_2017}. To correctly account for the UV photon flux at the top of the atmosphere, we use the same synthetic star spectra with coadded UV observations as mentioned in Section \ref{meth:temperature}. 

Some atmospheric regions may have dynamical timescales lower than chemical timescales, which effectively freezes the chemistry. Chemical abundances in these regions are controlled by the dynamics such as eddy or molecular diffusion.  Since in \textsc{VULCAN} K\textsubscript{zz} (Eddy diffusion) is a free parameter, we chose a commonly used value for rocky planets of 10\textsuperscript{8} cm\textsuperscript{2} s\textsuperscript{-1}. This value is constant all throughout the atmosphere. For molecular diffusion we calculate the coefficients specific for each species according to the dominant atmospheric constituent, which in most our cases is molecular nitrogen. However, molecular diffusion is only dominant at very low pressures which have insignificant contribution to the calculated spectra. For a fuller description regarding photochemistry, eddy diffusion and derivation of molecular diffusion coefficients as well as choices of other relevant parameters see \citep{ZILINSKAS_2020}.

\subsection{Synthetic emission spectra}
\label{meth:emission}

Finally, to simulate emission spectra we employ a radiative transfer code \textsc{petitRADTRANS}\footnote{http://gitlab.com/mauricemolli/petitRADTRANS} \citep{Molliere_2019}. As with \textsc{HELIOS}, we use a wavelength resolution of $\lambda/\Delta\lambda = 1000$. As input for opacities we used \ce{H2}, \ce{H2O}, \ce{HCN}, \ce{CN}, \ce{C2H2}, \ce{C2H4}, \ce{CO}, \ce{CO2}, \ce{CH4}, \ce{NH3}, \ce{OH}, \ce{NH}, \ce{CH}, \ce{NO}, as well as collision induced absorption of \ce{N2}-\ce{N2},\ce{N2}-\ce{H2} and \ce{H2}-\ce{H2}. For a more in depth description of opacities and how these are obtained see \citep{ZILINSKAS_2020,Molliere_2019}. It is convenient to represent the emission spectrum of a planet as planet-to-star contrast for which we, in order to assure continuity, use a generated PHOENIX \citep{Husser_2013} spectrum for a star of T\textsubscript{eff} = 5172 K, coinciding with the temperature of 55 Cancri.

\section{Results}
\label{sec:results}

\subsection{Temperature inversions}
\label{sec:TP}

\begin{figure*}
	\includegraphics[width=0.95\textwidth]{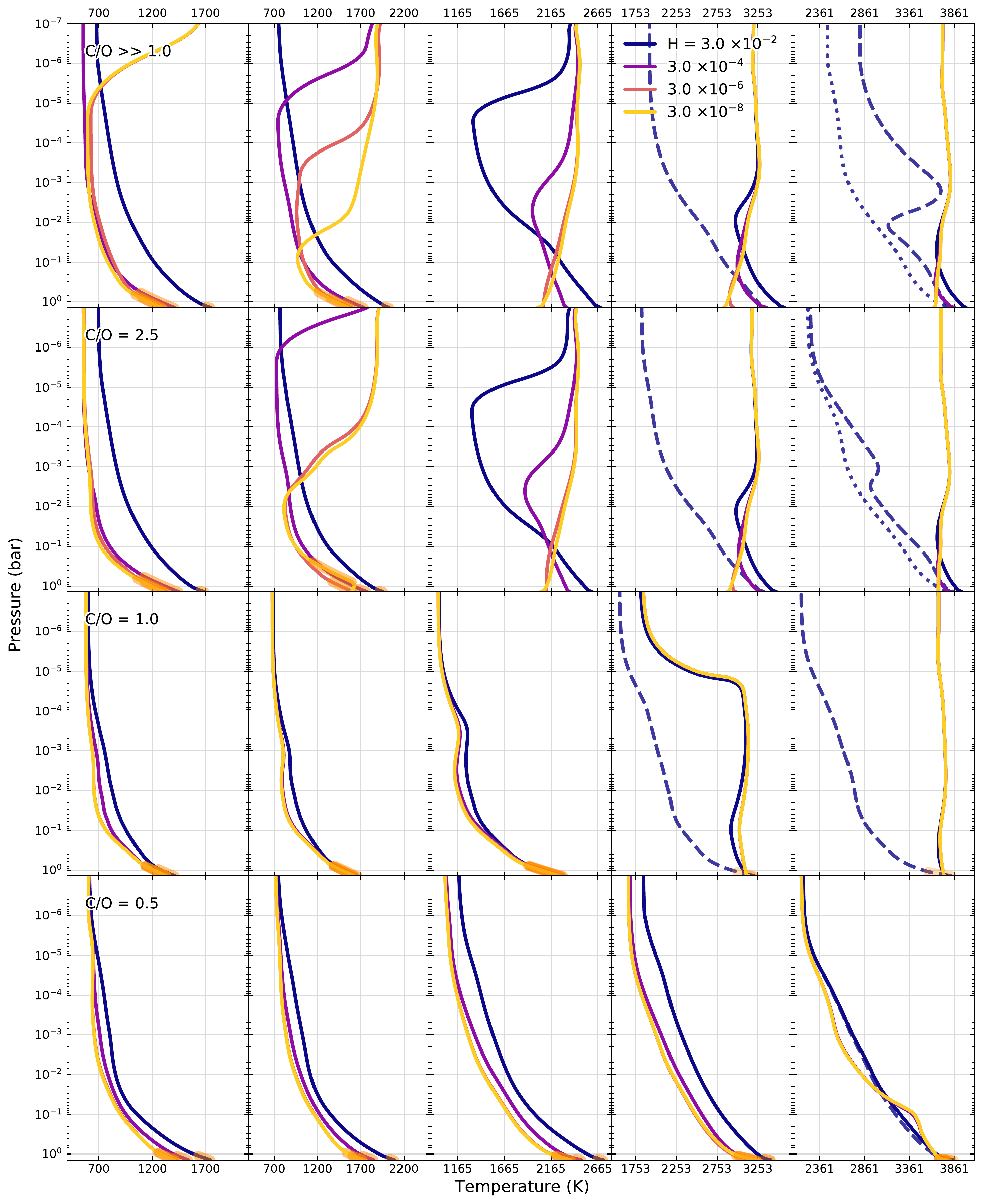}
    \caption{Temperature profiles in radiative-convective equilibrium computed using \textsc{HELIOS} for a super-Earth akin to 55 Cancri e. All cases are nitrogen-dominated and shown with varying mass fractions of hydrogen (See Table \ref{tab:comps} for exact compositions). Each row represents a different C/O ratio, each column - a different orbital range (0.06, 0.04, 0.02, 0.01, 0.007 AU from left to right), hence amount of irradiation that the atmosphere experiences. For each of the cases, the profiles are shown as a function of atmospheric pressure with P\textsubscript{surf} = 1.4 bar. Different colours indicate different initial abundances of hydrogen. Dashed curves show atmospheres where \ce{CN} opacity was omitted. Dotted curves indicate atmospheres where \ce{CN} and \ce{CH} opacities were removed. Some profiles have highlighted lower regions, where convection is the dominant energy transport mechanism. The overplotted horizontal grid represents a change of one order of magnitude in pressure, vertical grid indicates intervals of 500 K in temperature. The apparent temperature inversions are caused by shortwave absorption of \ce{CN} which abundance of increases with higher temperature and larger C/O ratios.}
    \label{fig:TP}
\end{figure*}

When comparing the thermal profiles of planets in the solar system, it has been noticed that in many cases the minimum temperature occurs at around 0.1 bar (boundary of troposphere). This lead to the development of an analytical approximation to model the temperature structure in these planets \citep{Robinson_2014}. It follows a dry adiabatic gradient from the surface to 0.1 bar with an isothermal structure for lower pressures. Recently this has been used in modelling rocky exoplanet atmospheres \citep{Morley_2017,Miguel_2019}. The approximation produces profiles similar to what we see in our lowest temperature C/O = 0.5 and C/O = 1.0 cases, shown in Figure \ref{fig:TP}. However, our simulations indicate that this estimation becomes less valid for hotter atmospheres and/or larger C/O ratios. In the following paragraphs we break down the behaviour of the temperature profiles for specific C/O cases.

\begin{figure}
	\includegraphics[width=\columnwidth]{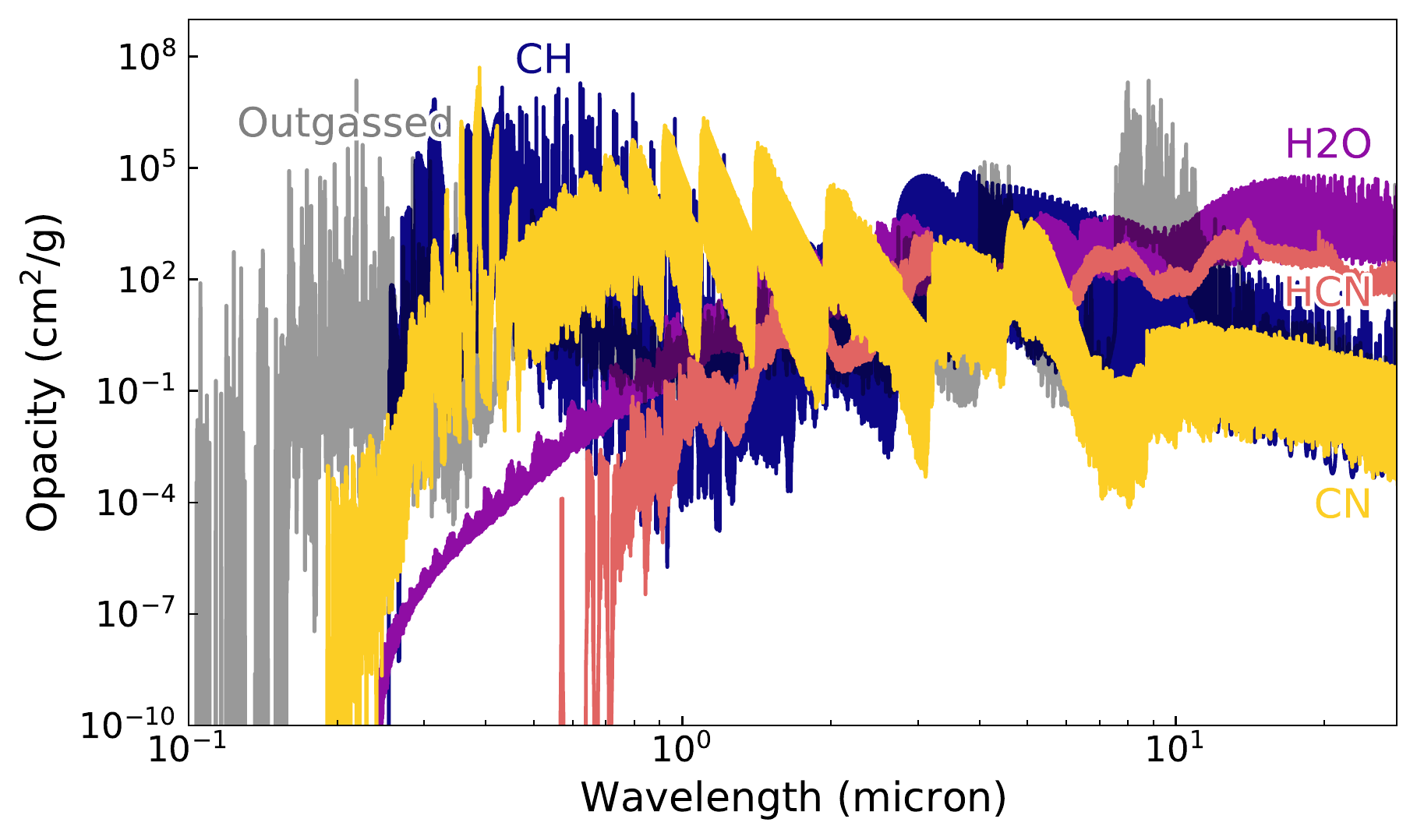}
    \caption{Opacity of CN in comparison with \ce{H2O}, \ce{HCN} and \ce{CH} for T = 2900 K and P = 1 bar. Also shown is the composite opacity of an outgassed 22 mbar magma ocean atmosphere with T = 3000 K (Zilinskas et al. in prep). For each species, the opacity is calculated assuming 100\% abundance and shown at a resolution of $\lambda/\Delta\lambda = 1000$.}
    \label{fig:opacities}
\end{figure}

For atmospheres with \textbf{C/O = 0.5} the chemistry is driven by the formation of oxides. The main sources of opacity are \ce{H2O} and \ce{CO2}, with \ce{CN} becoming abundant enough to contribute when T\textsubscript{Peff} = 3730 K. Moving from our lowest temperature cases to hotter, stronger irradiation causes a slight increase of abundances of the main opacity sources, pushing the photosphere upwards. The lower regions become optically thick, making them cool less efficiently, thus driving the surface temperature up. This, in turn, creates a larger difference in temperature between the bottom of the atmosphere (BOA) and the top of the atmosphere (TOA). The isothermal region is also slowly pushed up, which effectively removes the sharp troposphere boundary used in the analytical estimation. Reducing the hydrogen mass fraction results in decreased abundances of the main absorber \ce{H2O}, driving the photosphere closer to the surface, which allows for more efficient cooling. With C/O = 0.5, we only start to see slight effects of shortwave absorption at T\textsubscript{Peff} = 3730 K. The slight bump at 0.1 bar is caused by the accumulation of \ce{CN}, which is a strong shortwave opacity source. The absorption of incoming stellar irradiation by this molecule causes the atmosphere to heat up. Figure \ref{fig:opacities} displays the \ce{CN} opacity in comparison to \ce{H2O}, \ce{HCN}, \ce{CH} and a combined opacity of an outgassed metal-rich atmosphere from a magma ocean (Zilinskas et al. in prep). \ce{CN}, \ce{CH} and the metal-rich cases have strong shortwave absorption.

In the case of \textbf{C/O = 1.0} main opacity sources are \ce{CO}, \ce{H2O} and \ce{HCN}. If hydrogen is depleted, \ce{H2O} is replaced by \ce{CO2}, which along with \ce{CO} becomes the major opacity source. In the colder temperatures, these profiles exhibit similar structure to C/O = 0.5 models, however, once the atmosphere becomes hotter, the differences become substantial. At T\textsubscript{Peff} > 3000 K (4th column) many of the molecules (e.g. \ce{N2}, \ce{H2}, \ce{HCN}, \ce{CO}) are thermally dissociated, giving way for formation of \ce{CN}, which results in a strong temperature increase from the surface to $P= 10^{-5}$ bar, when compared to colder cases. This accumulation of \ce{CN} becomes greater as the temperature increases. The difference compared to models with no CN opacity (dashed) becomes very significant higher up in the atmosphere, however, it is less pronounced near the surface, where the photosphere lies. In the hottest cases, \ce{CN} opacity becomes strong enough to result in the profiles becoming near-isothermal.

Moving to \textbf{C/O > 1.0} cases, the results are very similar independent of how high the C/O ratio in the atmosphere becomes. When hydrogen is sufficiently abundant the main absorber is \ce{HCN}. With no hydrogen, \ce{CO} and \ce{CN} are the main sources of opacity. Low temperature models behave similarly to previously discussed ones with lower C/O ratios, however, C/O ratio above unity is now favourable for \ce{CN} formation, making inversions much more sensitive to the supply of hydrogen. If hydrogen is sufficiently depleted, the increase in \ce{CN} abundance causes strong temperature inversions, which can extend all the way to the planet's surface. In instances, the difference in temperature caused by inversions can be larger than 1000 K. For models where inversions extend to the surface (3rd, 4th and 5th columns) the temperature increase in the photosphere compared to the surface can be as high as 500 K. This has profound effect on the emission spectra (Section \ref{sec:emission}), effectively inverting absorption features. Even in models where hydrogen is abundant and inversions never extend all the way to the surface, emission spectra are still heavily affected. This is because the probed photospheric regions can reach pressures as low as P = $10^{-5}$ bar. In the figure the dashed curves indicate temperature profiles with \ce{CN} opacity removed, which causes inversions to vanish with an exception for the hottest cases, where \ce{CH} shortwave absorption becomes significant enough cause temperature inversions at pressures lower than $P= 10^{-2}$ bar. Dotted curves for these cases show that temperature profiles behave normally if, both, \ce{CH} and \ce{CN} opacities are removed.

\subsection{Effect of temperature inversions on atmospheric chemistry}
\label{results_chemistry}

\begin{figure*}
	\includegraphics[width=0.90\textwidth]{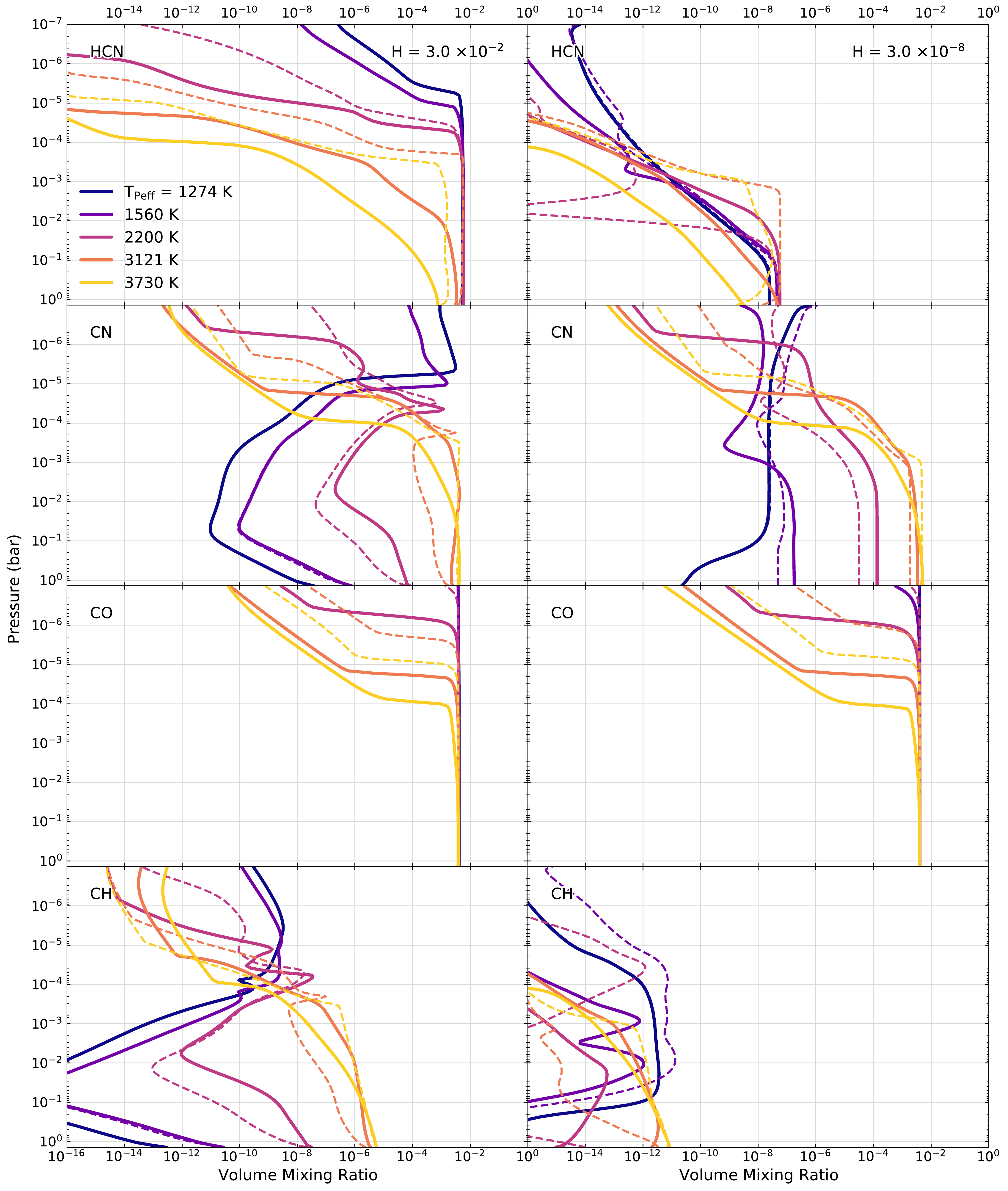}
    \caption{Volume mixing ratio of \ce{HCN}, \ce{CN}, \ce{CO} and \ce{CH} in nitrogen dominated hot super-Earth atmospheres. All cases shown for a C/O ratio of 2.5. Columns represent different hydrogen mass fractions, rows - different chemical compounds. The colour indicates the effective temperature of the planet. Solid lines represent atmospheres with temperature profiles shown in Figure \ref{fig:TP}, which in some cases have strong inversions. Dashed lines have \ce{CN} and \ce{CH} opacities omitted, hence lacking temperature inversions.}
    \label{fig:chemGrid}
\end{figure*}

The computed chemical abundances of the main species of interest (\ce{HCN}, \ce{CN}, \ce{CO}, \ce{CH}) are shown in Figure \ref{fig:chemGrid}. Focusing on temperature inversions, which are strongest for C/O > 1.0, we only show models with a C/O ratio of 2.5, where \ce{CN} formation is favourable.

In nitrogen atmospheres, \textbf{\ce{HCN}} formation is efficient when carbon is more abundant than oxygen \citep{ZILINSKAS_2020}, however, the resulting abundance is heavily dictated by the temperature and the supply of hydrogen. In cases with H = $3\times10^{-2}$ (left column), \ce{HCN} is the dominant carbon-nitrogen carrier for T\textsubscript{Peff} = 2200 K and below. For larger temperatures, \ce{HCN} is thermally dissociated, leaving \ce{CN} as the dominant carbon-nitrogen composite. With decreasing orbital radii photodissociation will also become increasingly more dominant, which along with thermal dissociation largely removes the presence of \ce{HCN} in the upper atmospheric regions. 

Since our atmospheres are nitrogen dominated with a relatively low amount of hydrogen in them, the abundance of \ce{HCN} is directly controlled by exactly how much \ce{H} there is. Reducing H from $10^{-2}$ to $10^{-8}$ directly results in the overall \ce{HCN} abundance being decreased by around 5 orders of magnitude, which is sufficient to push it below observability through low resolution emission (See Section \ref{sec:emission}).

Temperature inversions also affect the abundance of \ce{HCN}. The general effect is more clear with large temperatures and when H = $3\times10^{-2}$. The dashed models (no inversions) result in higher abundances of \ce{HCN} as it is less thermally dissociated. When T\textsubscript{Peff} = 1274 K or 1560 K, we do not see a large difference in abundance between inverted and non-inverted models as strong inversions do not occur here. The same is true for models with depleted hydrogen. Overall, where inversions are present, the increased thermal dissociation will result in \ce{HCN} abundances being lower.

Unlike \ce{HCN}, \textbf{\ce{CN}} abundance generally increases with higher temperatures. It also requires no hydrogen, which makes it consistently abundant between the two hydrogen cases. If \ce{HCN} is abundant, photochemical dissociation of it will directly increase the amount of \ce{CN} in the upper regions, as can be seen in the 3 coldest regimes with H = $3\times10^{-2}$. This effect is also confirmed with low hydrogen models, where much smaller abundance of \ce{HCN} does not result in photochemically enhanced abundances of \ce{CN}. Thus we suspect that in very hot atmospheres where \ce{HCN} can easily form, \ce{CN} would also show high abundances in the low pressure regions, which could have strong absorption (See Section \ref{sec:emission}).

With abundant hydrogen (left column), for the hottest cases (T\textsubscript{Peff} = 3121 K, 3730 K) \ce{CN} becomes the dominant carbon-nitrogen composite in the atmosphere. When \ce{H} is reduced (right column), \ce{CN} is almost always the dominant carbon-nitrogen carrier except for T\textsubscript{Peff} = 1274 K, where its production is slow.

Temperature inversions cause increased \ce{CN} formation with an exception in the T\textsubscript{Peff} = 3730 K model, where thermal dissociation becomes strong enough to reduce its abundance in the lower pressure regions, however, even in this case, the abundances near the surface are very similar to non-inverted atmospheres. Because inversions favour formation of \ce{CN}, the increased abundances would likely trigger even further temperature increase, which would yield more \ce{CN}, however, temperature profile iteration was not performed in this study.

In the shown atmospheres \textbf{\ce{CO}} is always the dominant carrier of oxygen. Generally, for C/O ratios exceeding unity, formation of other oxides, such as water or carbon dioxide, is restricted due to hydrogen being taken by \ce{HCN} and oxygen being in low supply in comparison to carbon. \ce{CO} abundance is also weakly affected by the changes in temperature, except for very low pressures, where it is increasingly dissociated with smaller orbits. Between the two hydrogen cases, the abundance of \ce{CO} is largely the same. Temperature inversions have no strong effect on its abundance, except for lowest pressure regions, where non-inverted atmospheres have slightly more \ce{CO}.

Since \textbf{\ce{CH}} is also a significant shortwave absorber, we show, that in cases with sufficient hydrogen and hot temperatures, its abundance rapidly rises to cause atmospheric inversions (See Section \ref{sec:TP}). Of course when hydrogen is depleted, \ce{CH} influence becomes vanishingly small. This indicates that in atmospheres with an abundance of carbon and hydrogen, \ce{CH} could potentially be a cause for strong atmospheric inversions, although its formation, much like of \ce{CN}, is confined to C/O ratios larger than unity and large temperatures.


\begin{figure}
	\includegraphics[width=\columnwidth]{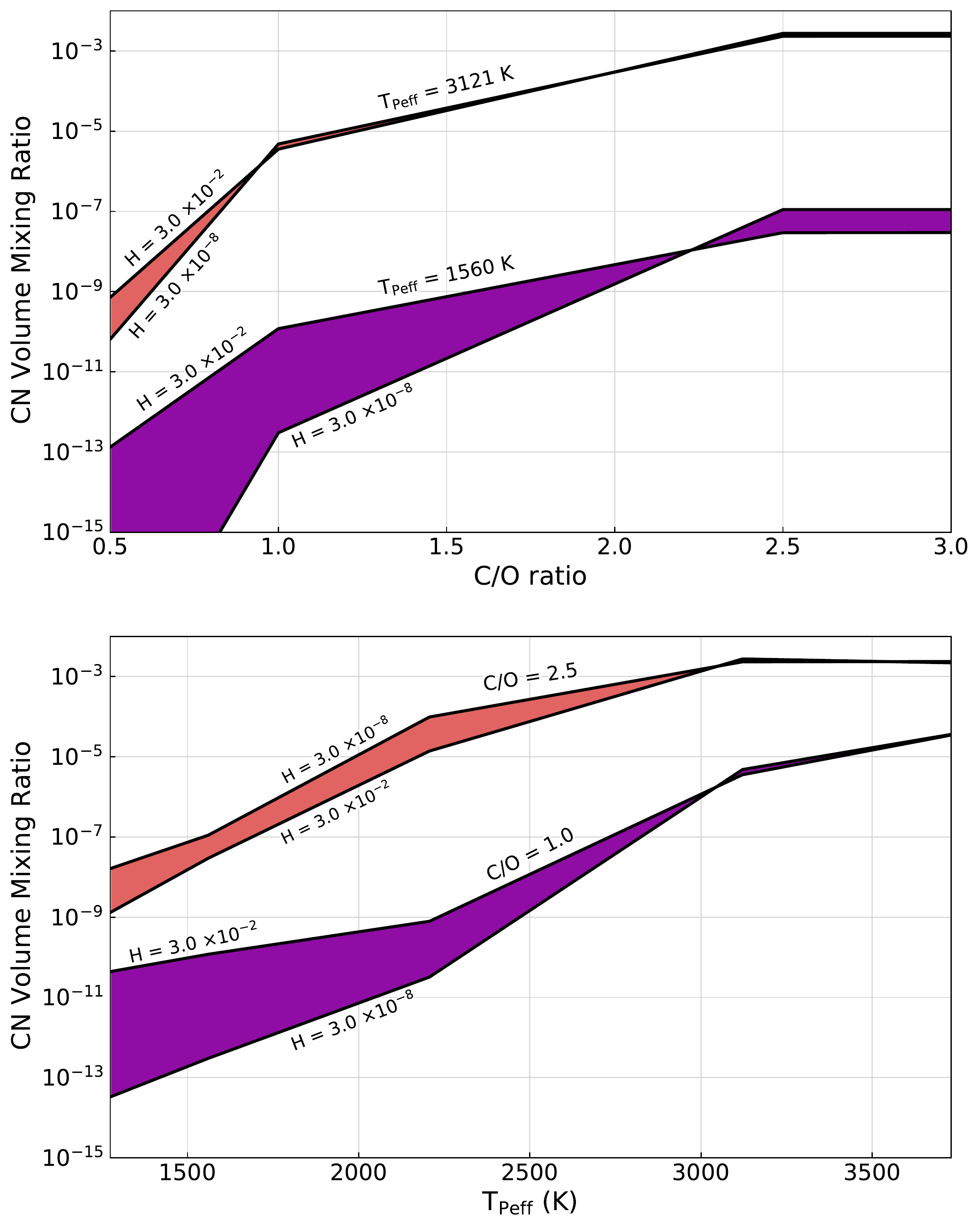}
    \caption{\ce{CN} volume mixing ratio as a function of C/O ratio (upper panel) and T\textsubscript{Peff} (lower panel). In the upper panel, different colours indicate different T\textsubscript{Peff}. In the lower panel, different colours indicate different C/O ratios. For both panels, the spread indicates the effect of initial hydrogen abundance. The mixing ratios are averaged values of the 1.4 - $10^{-4}$ bar pressure region, which mostly coincides with the usual photospheric emission region. The temperature profiles used in these atmospheres are shown in Figure \ref{fig:TP}.}
    \label{CO_CN}
\end{figure}
Figure \ref{CO_CN} shows the volume mixing ration dependence of \ce{CN} over different C/O ratios (upper panel) and different T\textsubscript{Peff} (lower panel). It is clear from the figure that \ce{CN} abundance is very sensitive not only to the C/O ratio, but also to temperature. With increasing temperature its volume mixing ratio becomes significantly higher. For the case of T\textsubscript{Peff} = 3121 K and C/O = 1.0 (top panel, red), its abundance is 5 orders of magnitude higher compared to T\textsubscript{Peff} = 1560 K of the same C/O ratio (top panel, purple), reaching a value near $10^{-5}$, which is sufficient to cause strong heating of the atmosphere. This highlights the importance of \ce{CN}, as its presence is not negligible even at C/O = 1.0, where \ce{CO} is thought to take up most of the carbon in the atmosphere. 

While the initial hydrogen abundance has an impact on \ce{CN}, its relative effects are minimal at sufficiently high temperatures and C/O ratios > 1.0. We notice that depending on the C/O ratio and temperature there are regions where less hydrogen results in less \ce{CN} being produced. This is not surprising since \ce{CN} formation is strongly correlated with the production of \ce{HCN}, which, at lower temperatures and lower C/O ratios forms when there is an overabundance of hydrogen compared to oxygen. Overall, hydrogen variance has small impact on the \ce{CN} abundance compared to parameters like temperature and C/O ratio.

\subsection{Emission via secondary eclipse}
\label{sec:emission}
Because most hot super-Earths are not expected to possess extended light-weight atmospheres, transmission spectroscopy is unlikely to be the proficient method of use for observations. Instead, due to their large day side temperatures, hot super-Earths are the ideal targets for planetary emission spectroscopy through observations of the secondary eclipse. That said, emission is entirely dependent on the probed temperature structure, hence inversions have large influence on the shape of the observed spectrum.

\begin{figure*}
	\includegraphics[width=0.9\textwidth]{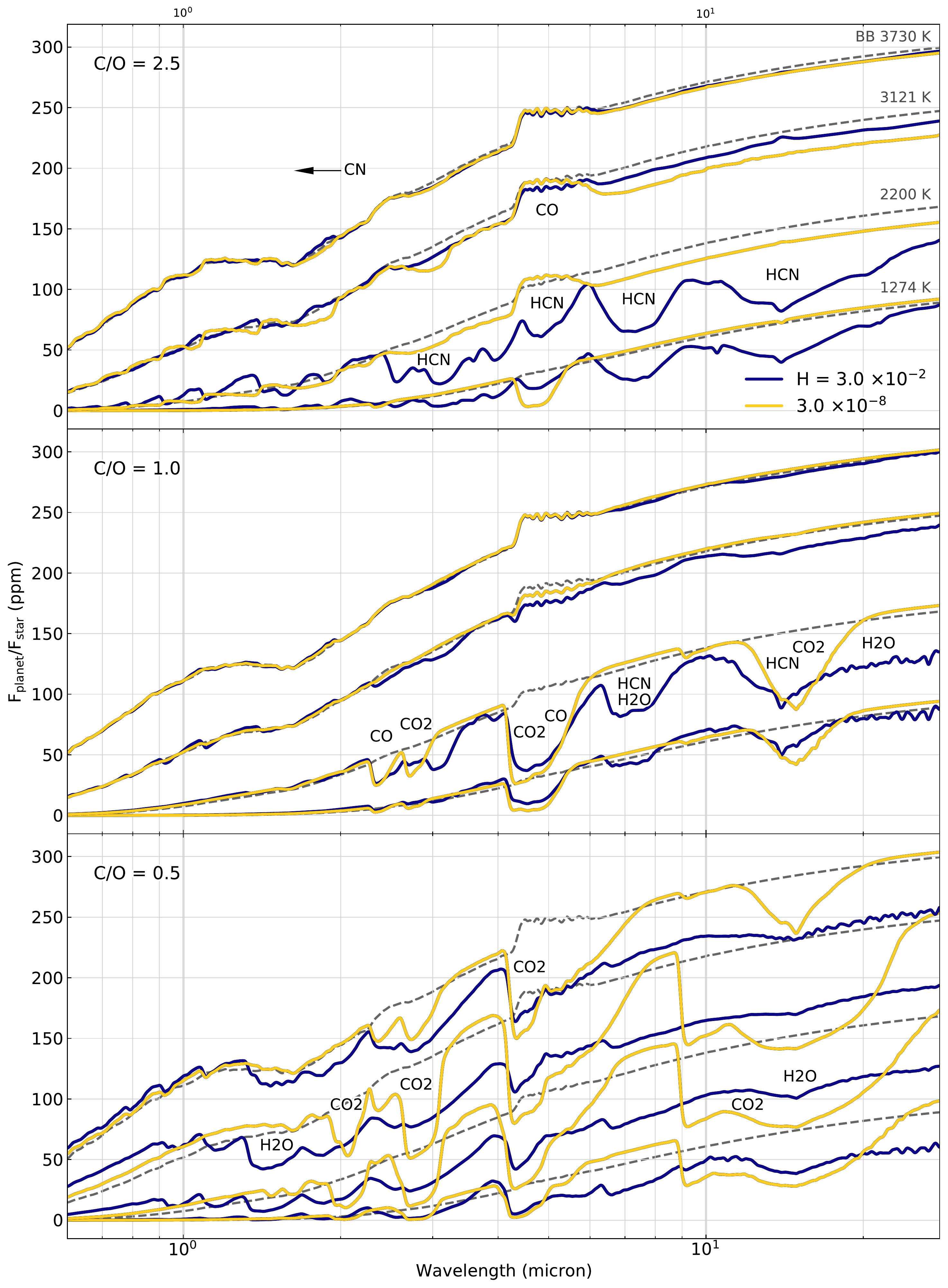}
    \caption{Synthetic emission spectra of a hot super-Earth with a nitrogen dominated atmosphere. The three panels represent three different C/O ratios (0.5 , 1.0 and 2.5). For each panel, 4 selected temperatures are shown, which coincide with the expected T\textsubscript{Peff} of the planet (T\textsubscript{Peff} of 1274, 2200, 3121 and 3730 K). The blackbody emission for each of the temperatures is indicated with dashed grey lines. The spectra are shown with two different hydrogen mass fractions, indicated by colour.}
    \label{emGrid}
\end{figure*}

In Figure \ref{emGrid} we show low resolution synthetic emission spectra of nitrogen dominated atmospheres for a range of temperatures with C/O ratios of 0.5, 1.0 and 2.5. To highlight differences, we show only the two extreme cases of hydrogen abundances. The used temperature profiles and compositions are described in Sections \ref{meth:temperature} and \ref{meth:chemistry}, respectively.

For \textbf{C/O = 0.5} oxygen is much more abundant than carbon, making formation of oxygen-rich molecules such as water and carbon dioxide efficient, which is mostly what we see regardless of the probed temperature range. Water is more dominant in the cases where the initial hydrogen abundance is high ($H = 10^{-2}$), and completely absent for cases when hydrogen is depleted, leaving \ce{CO2} as the major absorber.  Blue curves show absorption of mostly \ce{H2O} with \ce{CO2} at 4.5 \micron{} region. Yellow curves (representing atmospheres with low \ce{H}) are dominated purely by absorption of \ce{CO2}.  Such atmospheres do not have any shortwave absorbers and therefore we see no atmospheric inversions, regardless of the probed temperature range. Due to increased heating in the lower regions of the atmosphere, the continuum of the spectra tends to somewhat be higher than predicted blackbody emission (grey dashed).

Going to \textbf{C/O = 1.0} the chemistry changes, leaving most of the atmosphere in the form of carbon monoxide with some \ce{HCN}, \ce{H2O} or \ce{CO2} depending on the supply of hydrogen. For cases with T\textsubscript{Peff} of 1274 and 2200 K, we see strong absorption of \ce{CO} and \ce{CO2} at 4.5 \micron{}, with \ce{HCN} being present at 7 and 12 \micron{} when hydrogen is abundant. When hydrogen is depleted and the temperatures are low, instead of \ce{HCN} we see absorption of \ce{CO2} at 12 \micron{}.

For the 2 highest examined temperatures, T\textsubscript{Peff} = 3121  and 3730 K, large abundances of \ce{CN} cause inversion to occur, making the photospheric region largely isothermal, thus muting all the expected absorption features. Such spectra strongly coincide with the expected shown blackbody emission, regardless of whether hydrogen is depleted or not.

\begin{figure*}
	\includegraphics[width=0.85\textwidth]{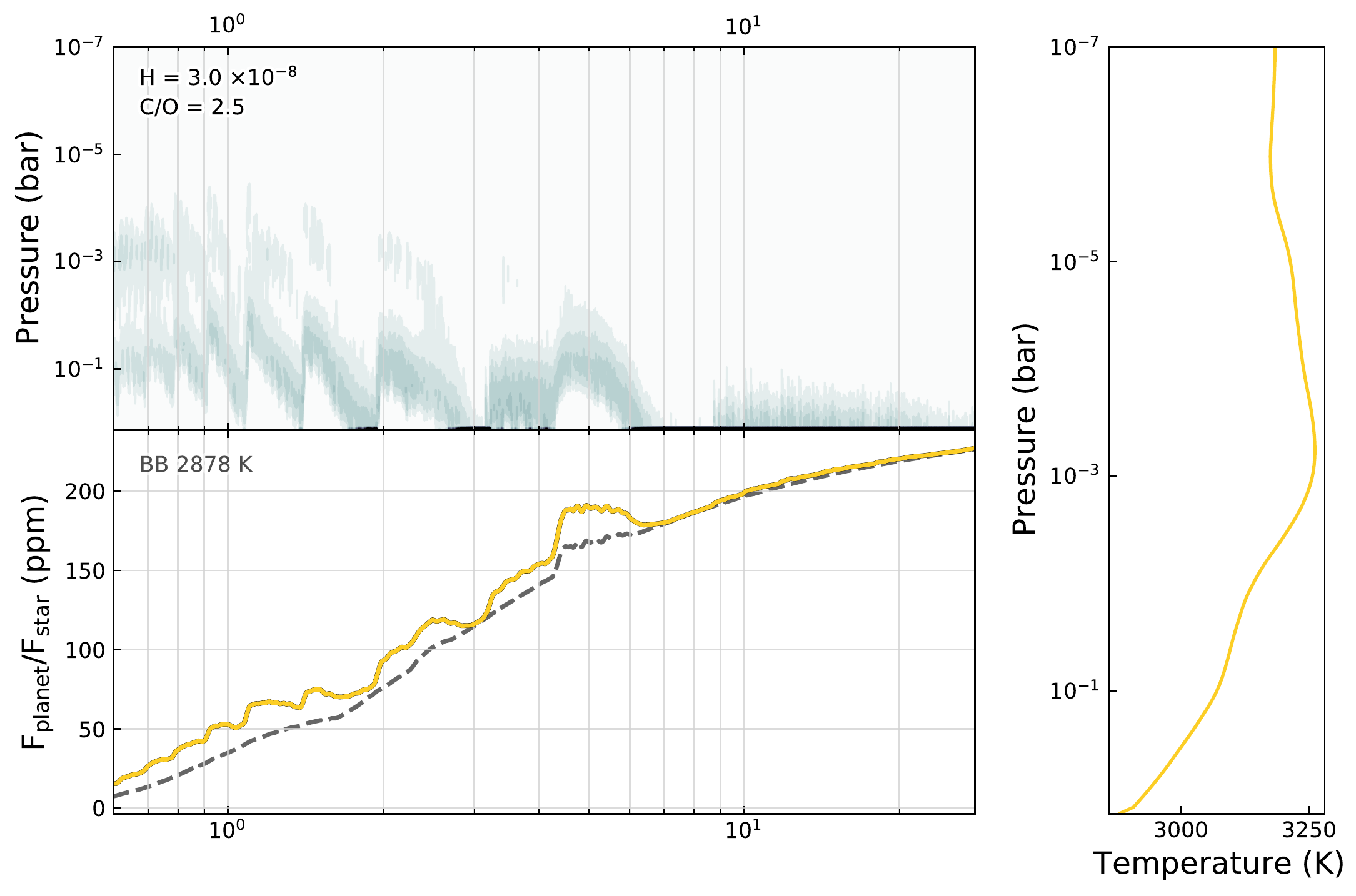}
    \caption{Contribution function of thermal emission of an atmosphere with an inverted thermal structure. The grey regions in the upper panel indicate the photosphere - optically thick regions being probed in emission. In the lower panel, indicated with yellow is a low resolution emission spectrum, with a coinciding surface temperature blackbody emission. The right panel indicates the corresponding temperature profile of the atmosphere.}
    \label{contribution}
\end{figure*}

Atmospheres with \textbf{C/O > 1.0} stop producing water and \ce{CO2}, making it mostly dominated by \ce{HCN} when hydrogen is in abundance or by \ce{CN} and \ce{CO} when hydrogen is depleted. For the cases of T\textsubscript{Peff} = 1274  and 2200 K, with hydrogen present, the absorption features are mostly from \ce{HCN} with \ce{CN} having minor ones below 2 \micron. If hydrogen is depleted, \ce{HCN} features vanish leaving \ce{CO} at 4.5 \micron{}, which in the case of 2200 K is shown as increased emission due to the now occurring temperature inversion. If hydrogen is sufficiently depleted, for these C/O ratios, we have temperature inversions even at 2000 K, see Figure \ref{fig:TP}. Going to higher temperatures, the inversions are present regardless of the hydrogen abundance, making the spectra look featureless in absorption, however, the continuum of the emission coming from the surface is much lower to what would be expected from a pure blackbody. This pushes the expected contrast of lower wavelengths to higher values. We illustrate this effect further in Figure \ref{contribution}.

Figure \ref{contribution} shows the emission contribution function for an extreme case with C/O = 2.5 and $H = 10^{-8}$. It is equivalent to the optically thick photosphere. The corresponding emission spectrum is indicated in the panel below it alongside blackbody emission with the planet's surface temperature. The right panel shows the full temperature profile. For wavelengths above 7 \micron{}, the corresponding emission region largely coincides with the planets surface, which, in this case is colder then the rest of the atmosphere. Going to lower wavelengths, instead of absorption features, we now see increased emission of \ce{CN} and \ce{CO} molecules. The emission is larger than predicted for a pure blackbody because the probed atmospheric regions have higher temperature. With increasing temperature \ce{CN} becomes increasingly more abundant, thus contributing from regions as high as $P = 10^{-5}$ bar, where even photochemistry plays a role in determining the molecular composition. While small temperature inversions can cause the atmosphere to show emission similar to a blackbody, the overall effect is significant as all of absorption features are removed.

\subsubsection{Effect of shortwave absorption on emission spectra}
\begin{figure*}
	\includegraphics[width=0.90\textwidth]{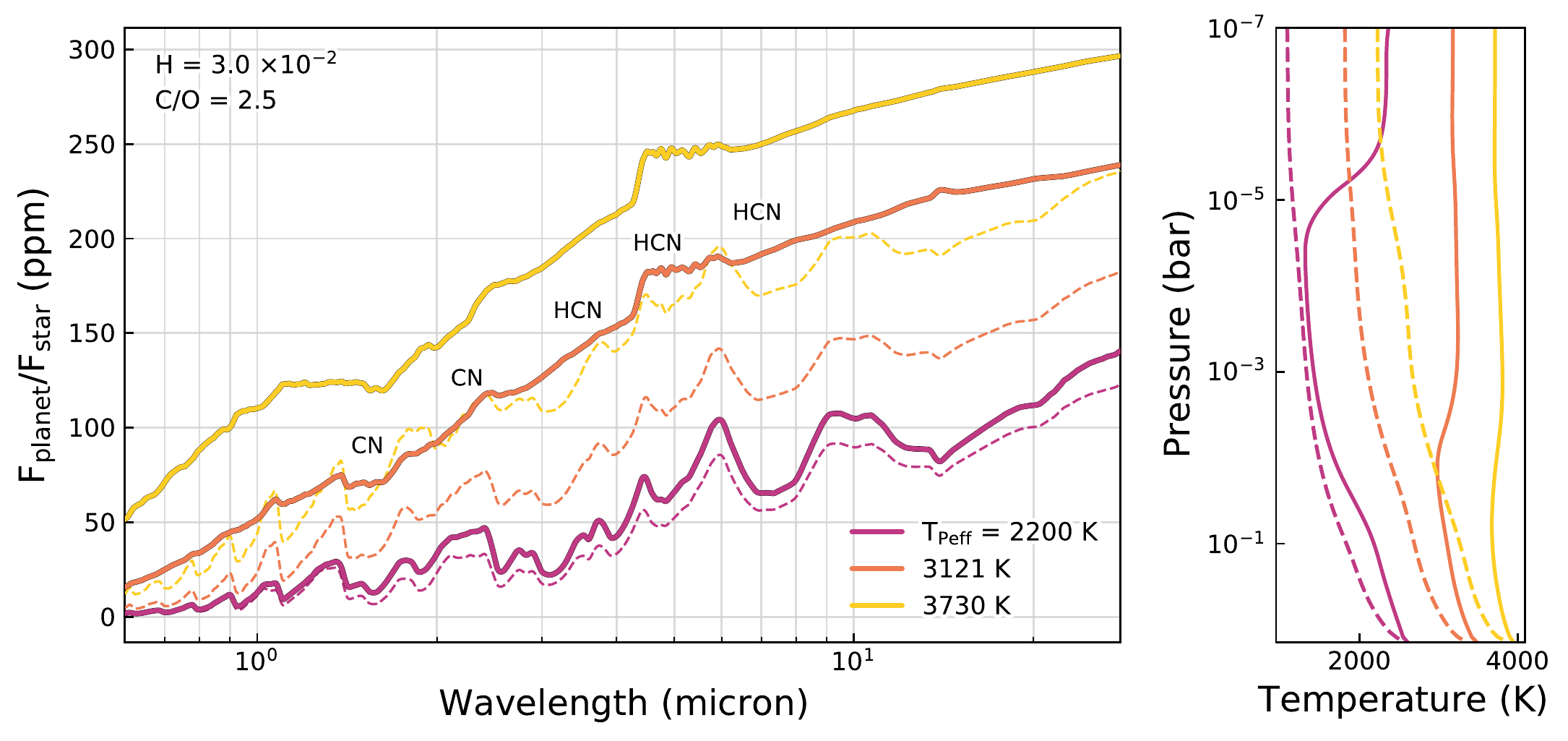}
    \caption{Comparison of emission spectra between atmospheres with strong shortwave absorbers (\ce{CN}, \ce{CH}, represented by solid curves) that cause temperature inversions and without (thin dashed lines). Shown for nitrogen dominated atmospheres with C/O = 2.5, which is favourable for formation of \ce{CN} (See Section \ref{meth:chemistry}). The compositions used have a hydrogen mass fraction of $3 \times 10^{-2}$. Below the selected temperature range, due to inefficient \ce{CN} formation, atmospheres are much less prone to temperature inversions and thus are not shown. The right panel indicates the corresponding temperature profiles.}
    \label{invHighH}
\end{figure*}

\begin{figure*}
	\includegraphics[width=0.90\textwidth]{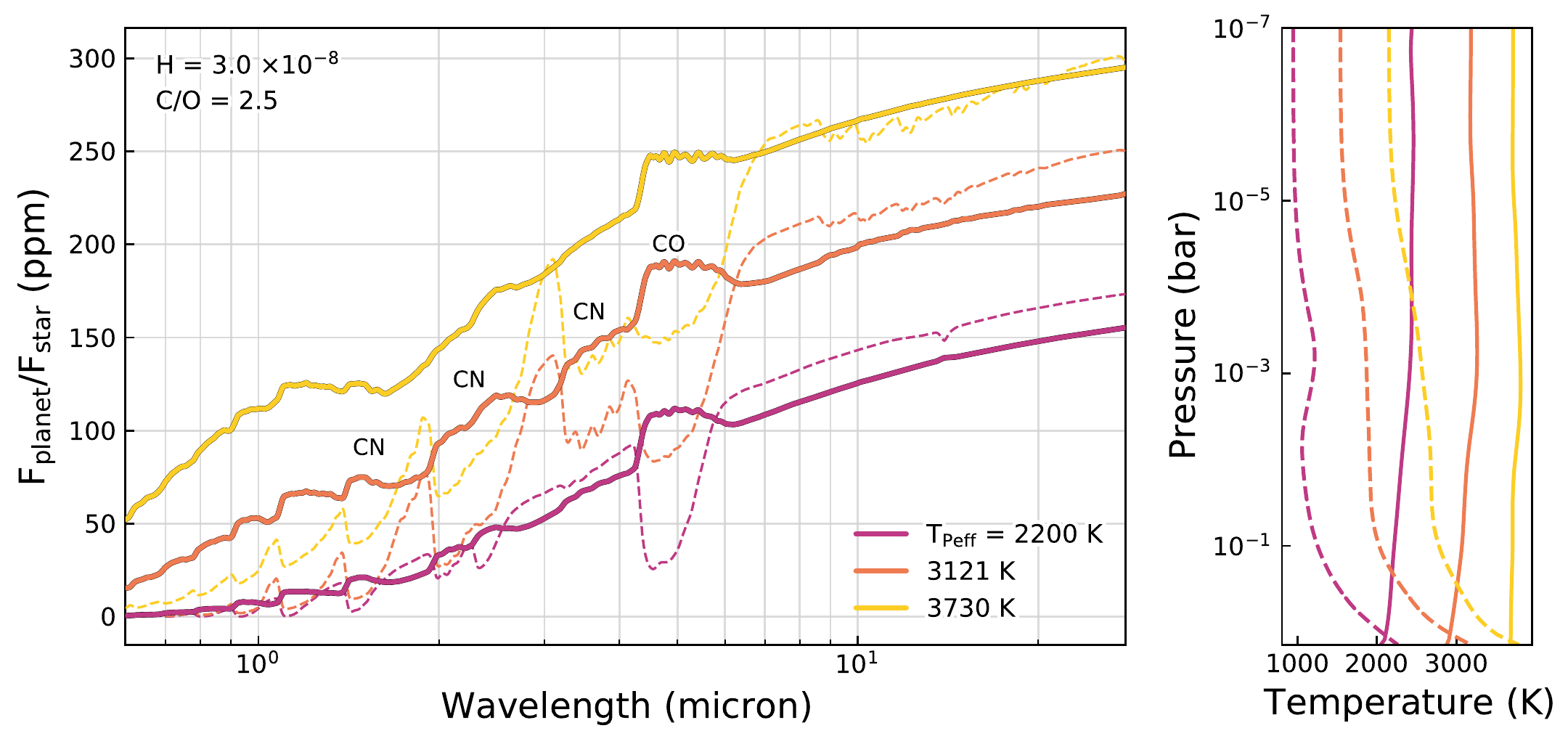}
    \caption{Emission spectra comparison between cases with temperature inversions and without. Same as in Fig. \ref{invHighH}, but with depleted hydrogen with a mass fraction of $3 \times 10^{-8}$.}
    \label{invLowH}
\end{figure*}

In Figures \ref{invHighH} and \ref{invLowH} we show a set of previously discussed atmospheres with temperature inversions (represented as solid curves) in comparison with calculated models where \ce{CN} and \ce{CH} shortwave absorption is removed (represented as dashed curves). The compositions used for these calculations are listed in Table \ref{tab:comps} and discussed in Section \ref{meth:chemistry}.

Removal of shortwave absorption removes temperature inversions, which drastically changes the predicted spectrum. Figure \ref{invHighH} represents models where hydrogen is abundant. As the temperature is increased, the inversions become stronger and the differences in the spectrum becomes larger. At T\textsubscript{Peff} of 2200 K, the inversion causes the photospheric region to be slightly hotter, however, the differences here are minor. Going to the case of T\textsubscript{Peff} of 3121 K, the jump is significant. The increased temperature from the occurring inversion causes emission to closely represent a blackbody, while the non-inverted case (dashed) emits from colder regions with absorption of \ce{HCN} and \ce{CN} throughout. We see the same behaviour if we go to an extreme case of T\textsubscript{Peff} of 3730 K.

The inversions become even more apparent when hydrogen is depleted. Figure \ref{invLowH} represents cases with $H = 3 \times 10^{-8}$. With no hydrogen, \ce{HCN} cannot efficiently form, thus leaving \ce{CN} and \ce{CO} as the largest opacity sources. Even at the lowest temperature case, the difference in \ce{CO} absorption at 4.5 \micron{} is nearing 100 ppm. Going to higher temperatures, \ce{CN} becomes increasingly abundant, making it a strong absorber at short wavelengths all the way to 5 \micron{}. The differences in emission between non-inverted and inverted cases here reach more than 100 ppm. \textit{This demonstrates how ignoring temperature inversions can result in vast misinterpretation of the expected atmospheric composition.}
\subsubsection{Difference in spectra between \ce{N2}, \ce{CO} and \ce{H2} dominated atmospheres}
\begin{figure*}
	\includegraphics[width=0.90\textwidth]{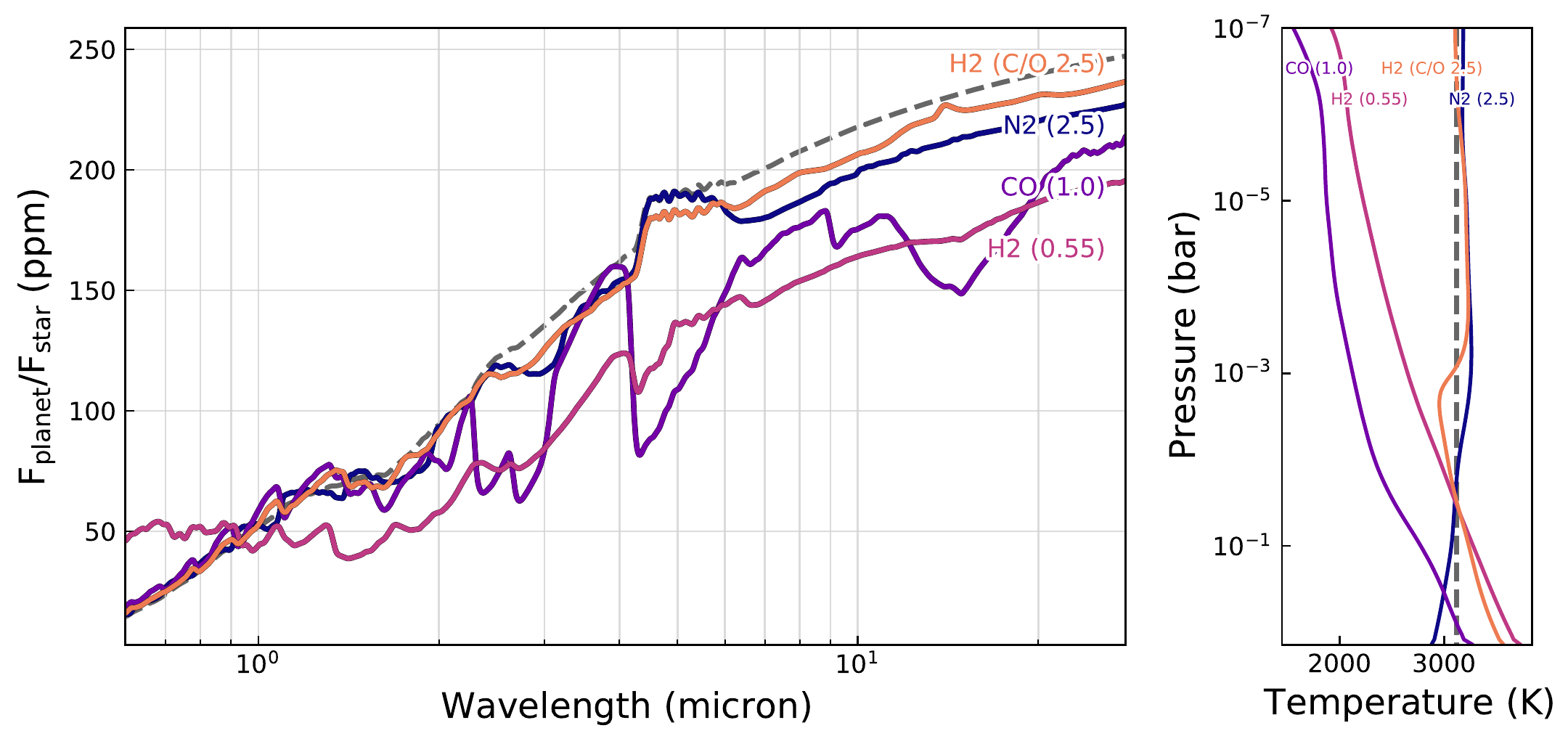}
    \caption{Comparison of synthetic emission spectra of atmospheres having different bulk compositions at T\textsubscript{Peff} = 3121 K. The indicated numbers in brackets represent atmospheric C/O ratio. Exact compositions are listed in the lower subsection of \ref{tab:comps}. Corresponding temperature profiles are displayed in the panel on the right.}
    \label{inverion}
\end{figure*}

So far we have shown that hot nitrogen atmospheres are prone to temperature inversions. In Figure \ref{inverion} we compare emission of an \ce{N2} atmosphere having an inverted thermal temperature structure with emission of model atmospheres dominated by other elements. These include: 96\% \ce{H2} atmosphere with C/O = 2.5, 96\% CO atmosphere with C/O = 1.0, and a 76\% \ce{H2} atmosphere with C/O = 0.55. The exact compositions are listed in the lower subsection of Table \ref{tab:comps}.

We find that for inversions to occur a nitrogen dominated composition is not necessary. A 96\% hydrogen atmosphere with a C/O ratio above unity still has inversions present, even with low amounts of carbon and nitrogen to form \ce{CN}. While the produced inversion does not reach the surface, the photospheric region coincides with pressures around $10^{-3}$ bar. The resulting emission spectrum is shaped by this part of the atmosphere, largely representing blackbody emission with parts having larger than expected flux. The overall continuum is slightly lower than the expected T\textsubscript{Peff} (indicated via a grey dashed line).

A 96\% \ce{CO} atmosphere with a C/O of 1.0 does not produce sufficient \ce{CN} to cause an inversion. All of the carbon is consumed to produce \ce{CO} and \ce{CO2}. However, this is not due to the C/O ratio being unity, but rather because there is insufficient amount of nitrogen in the atmosphere (0.7\%) to form large amounts of \ce{HCN} or \ce{CN}. As indicated in Figure \ref{fig:TP}, inversions do still occur when C/O = 1.0 if nitrogen is in large supply.

A hydrogen atmosphere with a C/O ratio much lower than unity, as indicated by \ce{H2} (0.55), is not prone to have thermal inversions. In such atmospheres no known shortwave absorbers form, thus producing the expected emission spectrum with deep absorption features from \ce{H2O} and \ce{CO2}

\section{Discussion}
\label{sec:discussion}


Our study is based on previous observations of 55 Cancri e, which lead us to believe that strongly irradiated super-Earths can have high-mean-molecular-weight atmospheres rich in nitrogen and/or carbon. While there is the outstanding issue of long-term day side emission variability on the planet \citep{Demory_2016a,Bourrier_2018}, the possibility of a nitrogen atmosphere is grounded on the fact that the majority of observations show little to no absorption through the 4.5 \micron{} \textit{Spitzer} IRAC band. As we have explained in this paper, the lack of the seen absorption features can be due to a temperature inversion and not necessarily due to the lack of certain molecular species in the atmosphere, as we suspect could be the case for 55 Cancri e. With the current secondary eclipse data, predicting the bulk composition of the planet's atmosphere is unrealistic. However, transmission HST observations of the atmosphere by \citet{Tsiaras_2016a} indicated possible absorption of \ce{HCN}, which, if true, would strongly point to an atmosphere with a C/O ratio above unity. Though the results are conflicting due to the predicted lack and rapid loss of hydrogen in the atmosphere \citep{Bourrier_2018}. It was also proposed by \citet{Fortney_2008} that temperature inversions would imply shorter radiative timescales, causing the planets to have a larger day-night temperature difference, which is exactly what we see on 55 Cancri e. Overall, strong irradiation, the lack of absorption at 4.5 \micron{}, the large day-night contrast and possible signs of \ce{HCN} are all, according to our models, in favour of thermal inversions occurring on the day side of the atmosphere.

More generally, our results indicate that irradiated atmospheres with T > 2000 K and C/O $\geq$ 1.0 are expected to host thermal inversions, even if the atmosphere is not dominated by nitrogen. Interestingly, this temperature threshold has been proposed as the requirement for hot Jupiters to have inversions \citep{Fortney_2008,Lothringer_2018}, now confirmed for multiple targets \citep{Haynes_2015,Evans_2017,Sheppard_2017}. Unlike in our study, these inversions occur not due to the presence of \ce{CN} but due to \ce{TiO}, \ce{VO} and other metals. In secondary atmospheres, these species are not likely to be substantially abundant when C/O is $\geq$ 1.0. Because of this, it was previously thought that atmospheres with C/O larger than unity could not have thermal inversions. However, we have shown that this is not the case if the atmosphere contains nitrogen and is sufficiently irradiated.

\ce{CN} is only one example of a molecule that is efficiently formed in hot atmospheres. It's abundance is enhanced by photodissociation and depletion of hydrogen, both of which are expected to take place for irradiated planets. In this paper we have explored only a simple chemistry case, neglecting the influence of the heavier metals being possibly outgassed from the surface magma, which, even at low abundances could have profound influence on the predicted temperature structure and emergent spectra. We also ignore ion chemistry and the effect of chemical kinetics on the produced temperature profiles. We aim to account for these effects in a more detailed, future study on outgassed atmospheres of hot super-Earths.

\section{Conclusion}
\label{sec:conclusion}

Using one-dimensional radiative transfer and chemistry models we have demonstrated that shortwave absorption of \ce{CN} and \ce{CH} can cause temperature inversions in strongly irradiated C/O $\geq$ 1.0 atmospheres of super-Earths, which we suspect could be the case for 55 Cancri e. \ce{CN} is one of the few molecules that we find to be extremely stable at large temperatures. It's abundance can be enhanced via photodissociation of other major species, such as \ce{CO} and \ce{HCN}, as well as hydrogen depletion. Such effects are predicted to be most extreme on commonly occurring short period super-Earths. Much like for hot Jupiters \citep{Fortney_2008,Lothringer_2018}, we find that only atmospheres with T > 2000 K are prone to thermal inversions.  

In this study we have explored a limited chemistry case, neglecting possible outgassed metal compounds from the surface magma ocean or photochemical effects on the produced temperature profiles. Due to an outgassed magma composition having strong shortwave opacity (as demonstrated in Figure \ref{fig:opacities}), we expect even the atmospheres that are devoid of volatiles to show temperature inversions.

With this we propose that temperature inversions on strongly irradiated super-Earths should be the expected norm. Because of their high occurrence rate, emission brightness and short orbital periods, hot super-Earths will prove to be some of the most extreme natural laboratories for theoretical predictions of atmospheric structure and chemistry. The day side of tidally locked super-Earths are also unlikely to posses clouds which often obscure results. This makes such planets perfect candidates for future observations. Accounting for possible temperature inversions and relevant chemistry may provide important constraints on atmospheric characterisation with the upcoming \textit{JWST} and \text{ARIEL} missions.

\section*{Acknowledgements}
We thank Matej Malik for insightful discussion on the topic of radiative transfer. We also thank the anonymous referee for their feedback, which helped us improve the quality of our manuscript. 

\section*{DATA AVAILABILITY}
The numerical results underlying this article were calculated using the open source codes \textsc{HELIOS}\footnote{https://github.com/exoclime/HELIOS}, \textsc{FastChem}\footnote{https://github.com/exoclime/FastChem}, VULCAN\footnote{https://github.com/exoclime/VULCAN} and \textsc{petitRADTRANS}\footnote{http://gitlab.com/mauricemolli/petitRADTRANS}. The opacity data were obtained from a public database\footnote{http://opacity.world}. Other relevant data will be shared on reasonable request to the corresponding author.




\bibliographystyle{mnras}
\bibliography{paper} 







\bsp	
\label{lastpage}
\end{document}